\documentclass[sigconf]{acmart}
\usepackage{hyperref}
\usepackage{subfigure}
\usepackage{enumitem}

\AtBeginDocument{%
  }

\copyrightyear{2025}
\acmYear{2025}
\setcopyright{cc}
\setcctype{by}
\acmConference[WWW Companion '25]{Companion Proceedings of the ACM Web Conference 2025}{April 28-May 2, 2025}{Sydney, NSW, Australia}
\acmBooktitle{Companion Proceedings of the ACM Web Conference 2025 (WWW Companion '25), April 28-May 2, 2025, Sydney, NSW, Australia}
\acmDOI{10.1145/3701716.3715192}
\acmISBN{979-8-4007-1331-6/25/04}
\begin{document}

\title{PhishIntel: Toward Practical Deployment of Reference-Based Phishing Detection}

\author{Yuexin Li}
\affiliation{%
  \institution{National University of Singapore}
  \country{Singapore}
}
\email{yuexinli@u.nus.edu}

\author{Hiok Kuek Tan}
\affiliation{%
  \institution{National University of Singapore}
  \country{Singapore}
}
\email{e1357001@u.nus.edu}

\author{Qiaoran Meng}
\affiliation{%
  \institution{National University of Singapore}
  \country{Singapore}
}
\email{qiaoran@nus.edu.sg}

\author{Mei Lin Lock}
\affiliation{%
  \institution{NCS Cyber Special Ops R\&D}
  \country{Singapore}
  }
\email{meilin.lock@ncs.com.sg}

\author{Tri Cao}
\affiliation{%
  \institution{National University of Singapore}
  \country{Singapore}
}
\email{tricao@nus.edu.sg}

\author{Shumin Deng}
\affiliation{%
  \institution{National University of Singapore}
  \country{Singapore}
}
\email{shumin@nus.edu.sg}

\author{Nay Oo}
\affiliation{%
  \institution{NCS Cyber Special Ops R\&D}
  \country{Singapore}
  }
\email{nay.oo@ncs.com.sg}

\author{Hoon Wei Lim}
\affiliation{%
  \institution{NCS Cyber Special Ops R\&D}
  \country{Singapore}
  }
\email{hoonwei.lim@ncs.com.sg}

\author{Bryan Hooi}
\affiliation{%
  \institution{National University of Singapore}
  \country{Singapore}
}
\email{dcsbhk@nus.edu.sg}

\renewcommand{\shortauthors}{Li et al.}

\begin{abstract}
  Phishing is a critical cyber threat, exploiting deceptive tactics to compromise victims and cause significant financial losses. While reference-based phishing detectors (RBPDs) have achieved notable advancements in detection accuracy, their real-world deployment is hindered by challenges such as high latency and inefficiency in URL analysis. To address these limitations, we present \textit{PhishIntel}, an end-to-end phishing detection system for real-world deployment. PhishIntel intelligently determines whether a URL can be processed immediately or not, segmenting the detection process into two distinct tasks: a fast task that checks against local blacklists and result cache, and a slow task that conducts online blacklist verification, URL crawling, and webpage analysis using an RBPD. This fast-slow task system architecture ensures low response latency while retaining the robust detection capabilities of RBPDs for zero-day phishing threats. Furthermore, we develop two downstream applications based on PhishIntel: a phishing intelligence platform and a phishing email detection plugin for Microsoft Outlook, demonstrating its practical efficacy and utility.
  
\end{abstract}






\maketitle

\section{Introduction}

The vast majority of cyber attacks originate from phishing\cite{cloudflare_report}. For instance, phishing attacks deceive users into giving away their credentials, such as via malicious emails with links to webpages that impersonate known brands. Such illegal campaigns have resulted in tremendous financial loss, establishing phishing as one of the most severe social engineering threats today. Consequently, deploying effective and efficient phishing detection systems is essential to safeguard network users from such harm.

Recent advancements in phishing detection have predominantly focused on enhancing detection effectiveness. Among these, reference-based phishing detectors (RBPDs) have garnered significant research attention \cite{phishpedia, phishintention, dynaphish, phishllm, knowphish}. These detectors rely on website content analysis, such as URLs, screenshots, and HTML, to determine phishing activity. Specifically, RBPDs classify phishing webpages by first determining the brand that the webpage appears to contain based on its screenshot and HTML, called its `brand intention'. If a webpage exhibits a certain brand intention but its domain fails to match any legitimate domains of that brand, it is classified as phishing. RBPDs rely on the fundamental invariant that attackers cannot create a webpage whose brand intention matches its actual domain: e.g., while attackers can mimic the PayPal brand, they cannot create a webpage with the actual PayPal domain (\url{www.paypal.com}). This enables RBPDs to achieve high precision and robustness while providing clear explanations for their predictions.

Despite their robustness and explainability, several challenges remain when deploying RBPDs in real-world scenarios, significantly affecting the user experience. \textbf{(1) Latency}. RBPDs fundamentally rely on webpage content for their analysis, but in an email setting, we only have direct access to URLs (which appear in the emails). While integrating a webpage crawler can address this limitation by fetching the required data, it significantly increases latency. We empirically find that loading a single webpage using a crawler can take up to ten seconds, creating substantial delays. Such prolonged wait times severely hinder user experience, particularly when handling a large volume of requests. \textbf{(2) Efficiency}. Processing all incoming URL requests indiscriminately can lead to unnecessary inference costs, especially for URLs that have already been publicly or locally analyzed. Therefore, a systematic and optimized phishing detection system architecture is needed.

To tackle these practical challenges, we introduce \textit{PhishIntel}, an end-to-end phishing detection system for real-world deployment. The key design principle of PhishIntel is to process the URLs with different tiers. Therefore, we design a fast-slow task system architecture. Specifically, the \textit{fast task} checks the URLs against publicly available blacklists and a result cache that stores the result of an RBPD. If a URL is in the blacklist or the cache, its phishing detection result can be immediately returned; otherwise, it will be sent to the slow task queue pending additional processing. The \textit{slow task}, on the other hand, handles tasks that take much longer processing time. This includes online blacklist verification, webpage crawling, and webpage content analysis through an RBPD. Once a result is obtained, it will be added back to the result cache for future reference, instead of analyzing it again.

The complete PhishIntel system design ensures low response latency while retaining the robust detection capabilities of RBPDs for zero-day phishing threats. To validate its practical utility and effectiveness, we demonstrate PhishIntel with a phishing intelligence platform and a phishing email detection plugin in an enterprise-level email system. We also provide a case study to understand the effect of major system components. In summary, we make the following contribution:
\begin{itemize}[leftmargin=0.33cm, itemindent=0cm]
    \item \textbf{Phishing Detection System}. We propose PhishIntel, an end-to-end phishing detection system featuring real-time response while retaining the ability to detect zero-day phishing. 
    \item \textbf{Practical Applications}. We develop two applications to demonstrate the practical utility of PhishIntel: (1) a phishing intelligent platform to analyze phishing URLs and track emerging phishing trends, and (2) a phishing email detection plugin integrated into Microsoft Outlook for safeguarding users from email phishing.
\end{itemize}

\section{Methodology}
\subsection{Problem Statement}
Phishing detection is essentially a binary classification task. Given a URL $u$, a phishing detector $\mathcal{D}$ analyzes the information associated with $u$, such as its webpage screenshot $s$ and HTML $h$, and outputs a detection result $r$, classifying it as either benign or phishing.

In real-world scenarios, a deployed phishing detection system $\mathcal{F}$ typically processes a large volume of URLs. Formally, each incoming batch of URLs can be represented as a list $\mathcal{U}=\{u_1, u_2, ..., u_n\}$, with the corresponding detection results denoted as $\mathcal{R}=\{r_1, r_2, ..., r_n\}$, where $n$ is the number of URLs in this batch. Their associated webpage screenshots $\mathcal{S}=\{s_1, s_2, ..., s_n\}$ and HTML $\mathcal{H}=\{h_1, h_2, ..., h_n\}$ are not initially available and must be fetched using an additional webpage crawler $\mathcal{C}$. However, integrating $\mathcal{C}$ introduces significant runtime overhead, leading to increased system latency. This latency is particularly problematic in real-world applications, such as phishing email detection, where high throughput and responsiveness are critical for a positive user experience.

To address this challenge, we propose a fast-slow task system architecture. This approach categorizes URLs into those that can be processed instantly, providing immediate results, and those requiring further analysis, thereby balancing efficiency and zero-day phishing detection ability in large-scale URL streams.

\subsection{Overview}

\begin{figure}[t]
    \centering
    \includegraphics[width=\linewidth]{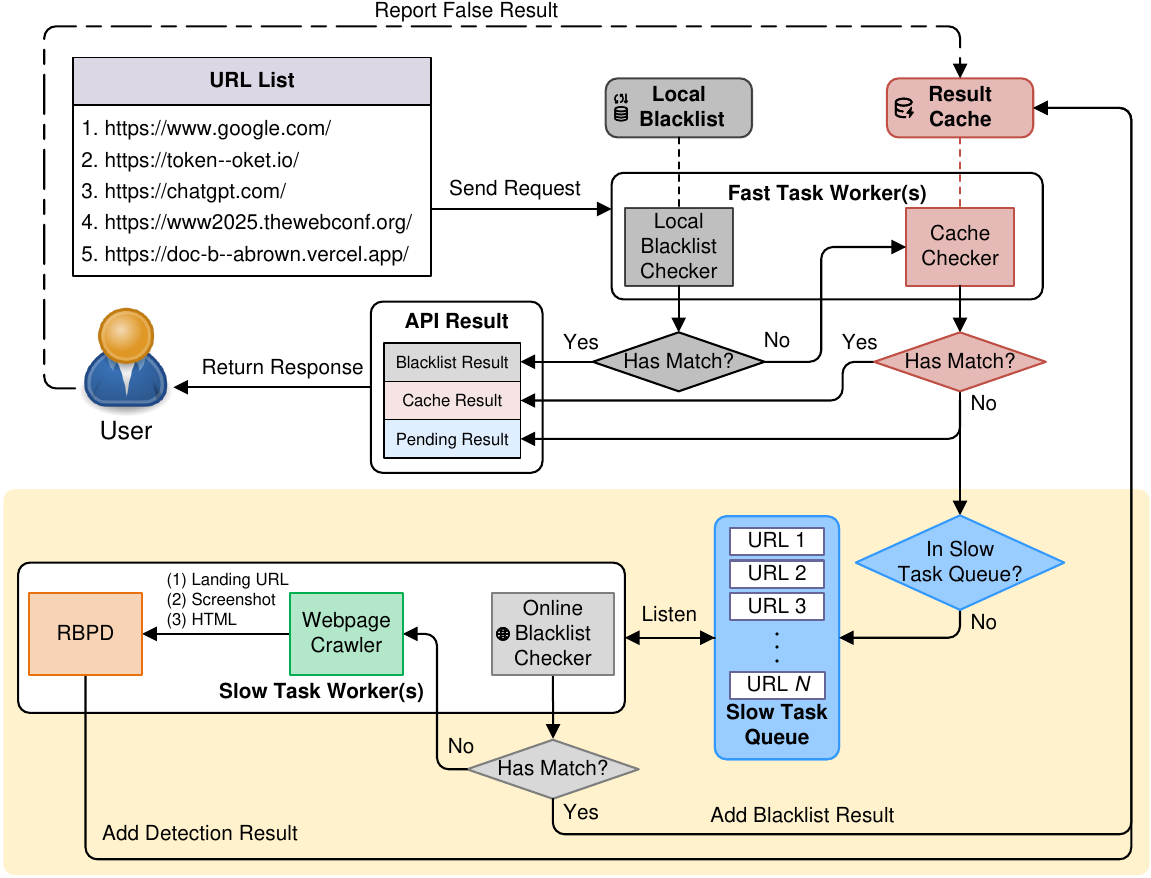}
    \caption{An overview of PhishIntel: URLs are first processed through FTW, with those failing to obtain results forwarding to the slow task queue for pending STW analysis.}
    \label{fig:phishintel}
\end{figure}

\begin{figure*}[htbp]
    \centering
    \includegraphics[width=1\linewidth]{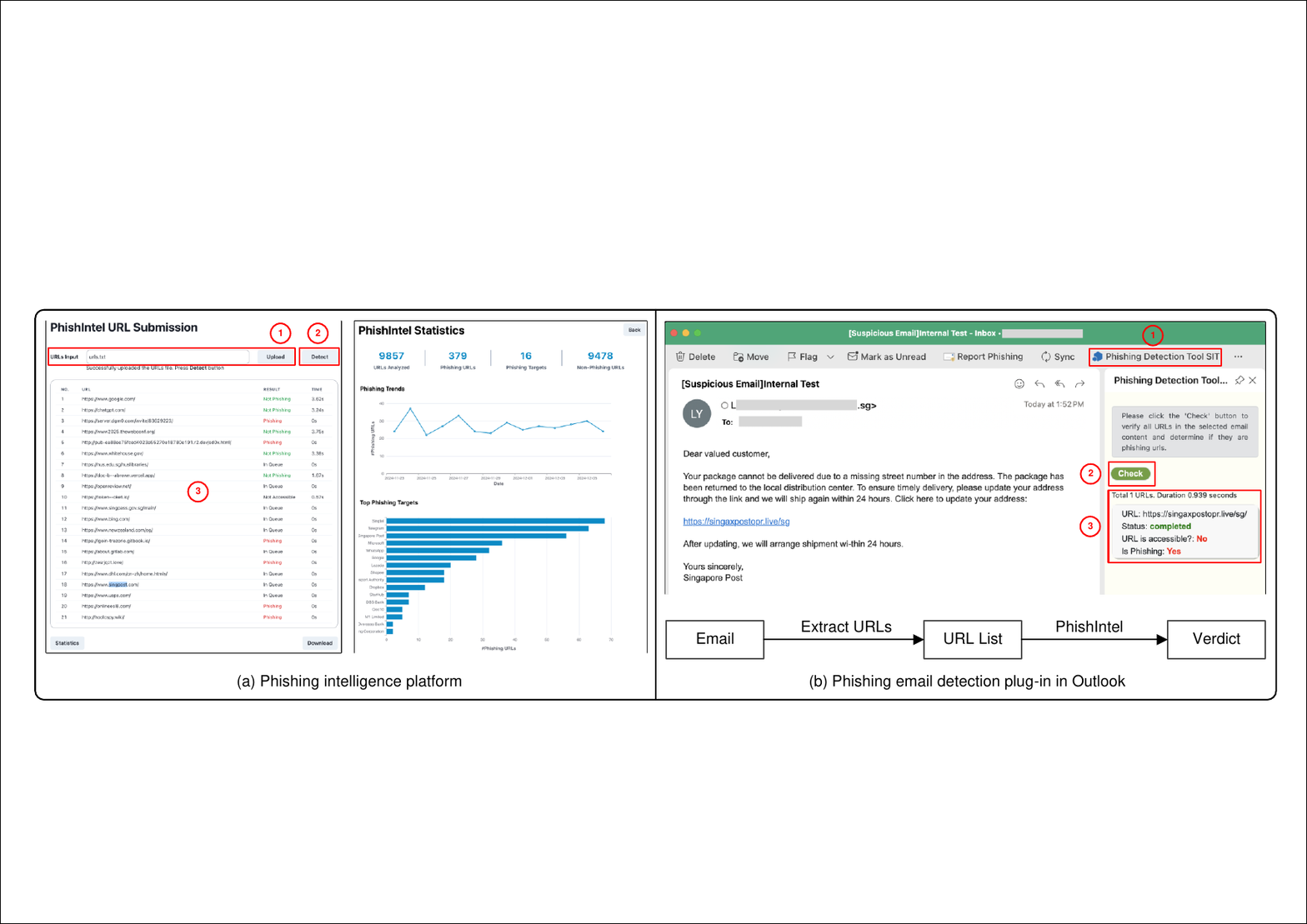}
    \caption{Illustration of two downstream applications built upon PhishIntel: (a) is a phishing intelligence platform, and (b) is a phishing email detection plugin in Outlook.}
    \label{fig:demo}
\end{figure*}

\hyperref[fig:phishintel]{Figure \ref{fig:phishintel}} offers an overview of \textit{PhishIntel}, our proposed phishing detection system designed with a fast-slow task architecture. The system is composed of three primary components: (1) the \textbf{Fast Task Worker (FTW)}, which processes incoming URL lists and promptly returns preliminary analysis results; (2) a \textbf{Slow Task Queue}, which temporarily stores URLs that the FTW cannot handle for more comprehensive analysis; and (3) the \textbf{Slow Task Worker (STW)}, which retrieves URLs from the Slow Task Queue to conduct in-depth phishing analysis.

\subsection{Fast Task Worker}
End-to-end phishing URL detection using state-of-the-art RBPDs is often a time-consuming task\cite{phishpedia, phishintention, dynaphish, knowphish, phishllm}, particularly when integrated with a webpage crawler to obtain necessary webpage data for analysis. For instance, we empirically observe that the webpage crawler can take up to 10 seconds to fetch webpage content, and a recent RBPD can spend another 10 seconds analyzing the webpage content\cite{dynaphish, knowphish}.  In contrast, users in real-world scenarios generally expect a response within a reasonable timeframe—typically around 2 seconds. Therefore, a straightforward combination of a crawler and an RBPD fails to meet the practical requirements for deployment.

Our FTW is designed to mitigate latency issues by enhancing the system efficiency. An FTW is composed of two components: a local blacklist checker and a result cache checker. These components enable the system to efficiently categorize URLs into those with immediately available results and those requiring further analysis. Specifically, if a URL $u$ is found in the local blacklist, the FTW will classify it as phishing and promptly return the result. If $u$ is not present in the local blacklist but matches an entry in the result cache, the FTW directly retrieves and returns the cached result. Conversely, if $u$ fails both checks, the FTW returns a `pending' result, indicating that $u$ will be forwarded to the slow task queue for further analysis. All these URL-matching processes are runtime-efficient, ensuring rapid processing and minimizing response latency.

The local blacklist is initialized using publicly available URL blacklists, effectively eliminating redundant analyses of known phishing URLs. To maintain its accuracy and relevance, a dedicated service process periodically updates the local blacklist by synchronizing with the public blacklist provider. The result cache is continuously updated with outputs from the slow task worker, which will be discussed in detail in the subsequent section. We additionally provide a means for users to provide their feedback on any detection results, to correct missing or inaccurate results. User feedback will be manually checked and the corrected entries will be placed in the cache.

\subsection{Slow Task Worker}
The STW is responsible for performing additional analysis on the URLs in the slow task queue, which often requires significantly more processing time than the FTW does. It operates by continuously listening to the slow task queue. Whenever the STW is available, it will fetch a URL from the slow task queue to perform analysis, if the slow task queue is not empty.

Specifically, an STW consists of three major components, an online blacklist checker, a webpage crawler, and an RBPD. The online blacklist checker initially checks $u$ against proprietary, non-publicly accessible blacklists via their URL matching APIs. This serves as an additional layer of the URL filtering process to complement FTW. If a match is found, the STW records a phishing result in the result cache. Otherwise, the webpage crawler retrieves the webpage content of $u$, including its screenshot and HTML. This content is subsequently analyzed by the RBPD, which generates a detection result that is also stored in the result cache.

\subsection{Implementation Details}
We build the FTW with Quart\footnote{https://github.com/pallets/quart} and Gunicorn\footnote{https://gunicorn.org/}. The local blacklist is instantiated with the PhishTank database\footnote{https://phishtank.org/}, while the result cache is implemented using Redis\footnote{https://redis.io/}. The STW and slow task queue are built with Celery\footnote{https://github.com/celery/celery}. We use Google Web Risk \footnote{https://cloud.google.com/security/products/web-risk} as the online blacklist checker, Playwright\footnote{https://playwright.dev/python/} as the webpage crawler, and the KnowPhish Detector\cite{knowphish} as the RBPD.

To optimize resource utilization and enhance request throughput, PhishIntel is deployed with 8 FTWs and 4 STWs. The complete system operates on an AWS EC2 instance with an NVIDIA L4 Tensor Core 24GB GPU.

\section{Demonstrations}

We demonstrate the utility of PhishIntel via two applications: (1) phishing intelligence platform, and (2) phishing email detection, as shown in \hyperref[fig:demo]{Figure \ref{fig:demo}}.

\subsection{Phishing Intelligence Platform}

The first application we developed is a web-based phishing intelligence platform that primarily offers two core functionalities: phishing URL analysis and a phishing statistics dashboard. Users can conduct phishing detection on a list of URLs by uploading a text file containing the URLs through the \texttt{Upload} button on the webpage. After uploading, users can click the \texttt{Detect} button to retrieve phishing detection results for the provided URLs. \hyperref[fig:demo]{Figure \ref{fig:demo}}(a) illustrates the phishing analysis results for 21 URLs. These results are directly returned by the FTWs, leveraging the local blacklist and result cache. URLs requiring further analysis are labeled as \texttt{In Queue}, indicating that they are still pending the STW processing.

This platform also features a dashboard page that visualizes phishing detection statistics derived from the result cache in PhishIntel, offering deeper insights into the current phishing attack landscape. For example, users can explore the number of unique phishing URLs, targeted entities, and recent trends, such as the daily detection count of phishing URLs and the most frequently targeted entities. Deployers can utilize this data to notify users about emerging phishing trends, safeguarding them from such attacks through regular reminders and alerts.

\subsection{Phishing Email Detection}
We also integrate PhishIntel into an enterprise-level email system, as a phishing email detection plugin in Outlook. Specifically, we create a button in the tool menu. When users encounter a suspicious email and would like to report it for analysis, they can click this button to activate the phishing email detection tool. After clicking the \texttt{Check} button in the tool, it will automatically extract all the URLs in the selected email as a list, and send it to PhishIntel. For the particular case in \hyperref[fig:demo]{Figure \ref{fig:demo}(b)}, the only URL in the email is classified as phishing, hence, the email is deemed as a phishing email.

\subsection{Performance Evaluation}

\begin{figure}
    \centering

    \begin{minipage}{0.495\linewidth}
        \centering
        \flushbottom
        \subfigure[Comparison of the average response time with different system architectures.]{
        \includegraphics[width=0.9\linewidth]{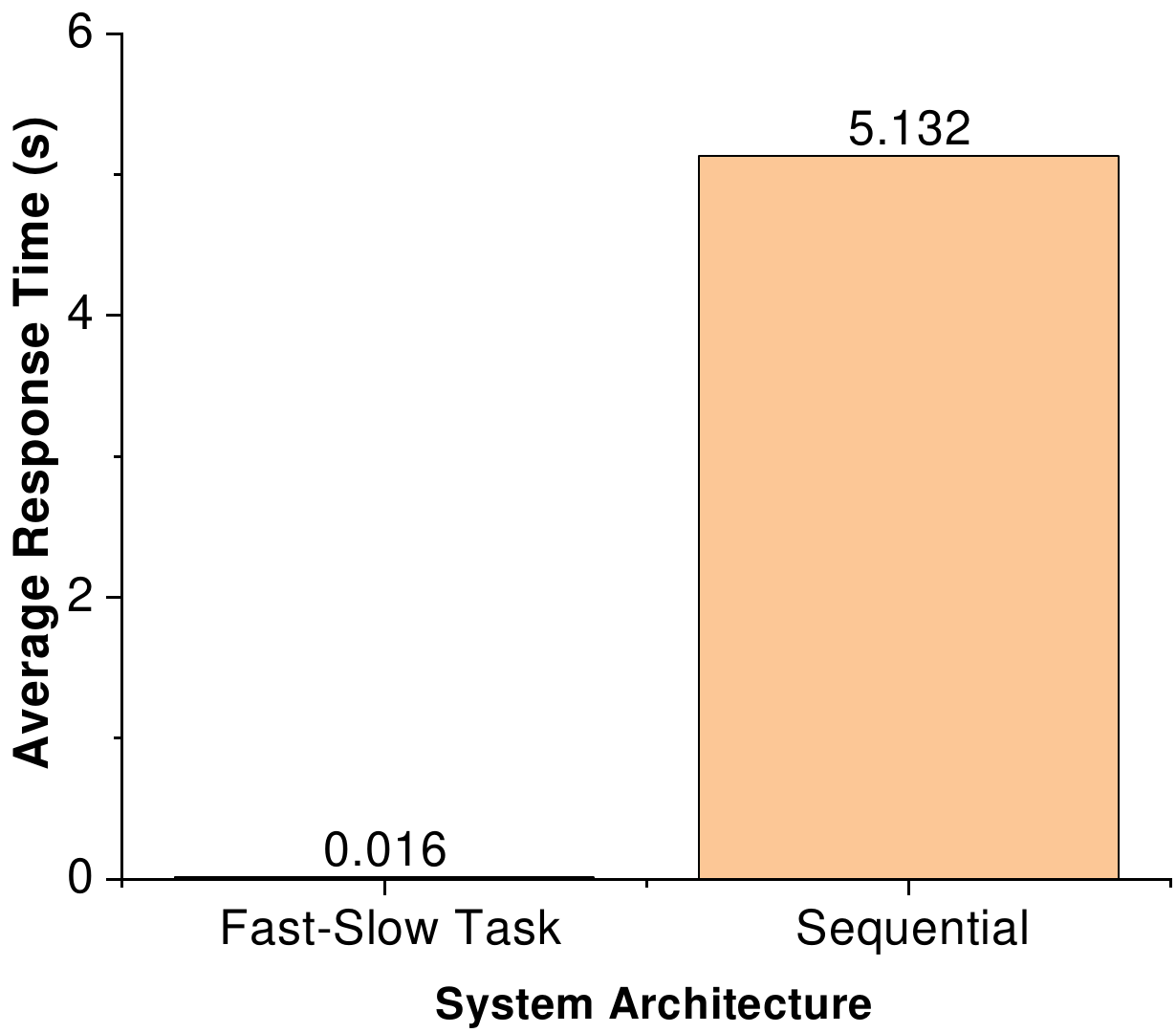}
        }
        \label{fig:evaluation_runtime}
    \end{minipage}
    \begin{minipage}{0.495\linewidth}
        \centering
        \flushbottom
        \subfigure[Distribution of the phishing reports from blacklist and RBPD.]{
        \includegraphics[width=0.9\linewidth]{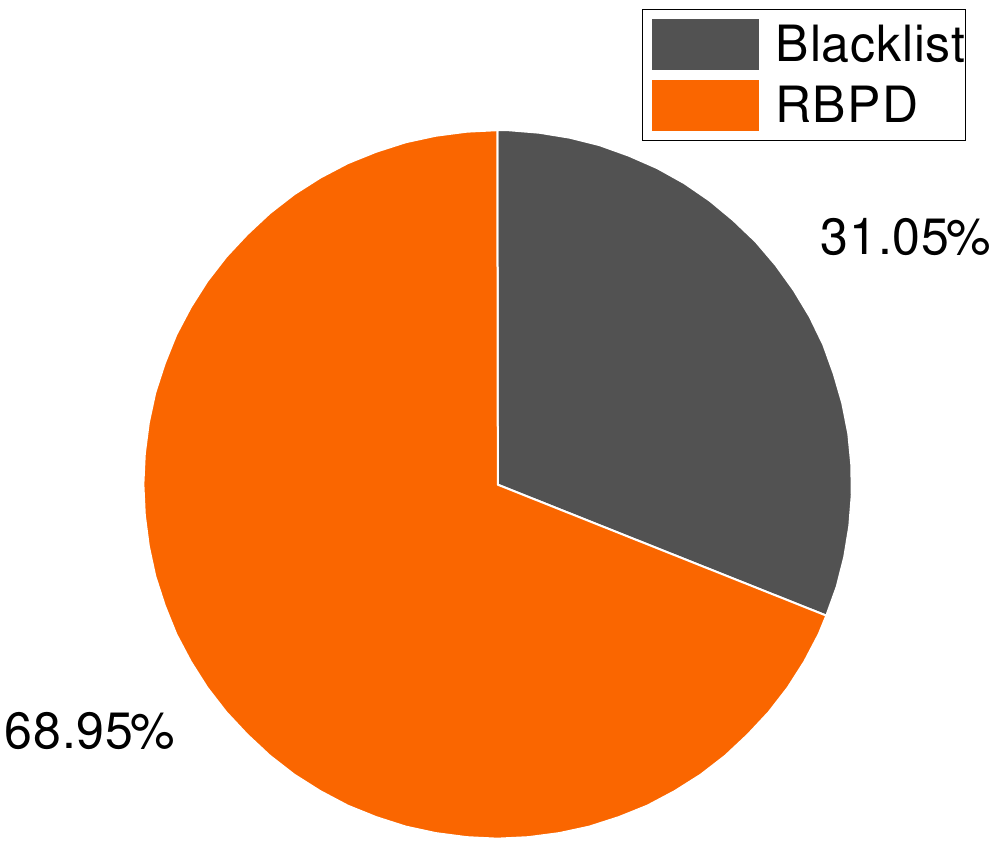}
        }
        \label{fig:evaluation_url_distribution}
    \end{minipage}
    \caption{Performance evaluation of PhishIntel.}
    \label{fig:evaluation}
\end{figure}

We further evaluate two key performance indicators of PhishIntel: the system response latency and the impact of blacklist filtering. For response latency, we compare the average response times of two architectural approaches: a fast-slow task design and a simple sequential pipeline, where the latter generates responses only after all components have completed their respective analyses. For blacklist filtering, we measure the proportion of phishing reports derived from blacklists and the RBPD. To conduct this evaluation, we input 1k randomly selected benign URLs from Tranco\footnote{https://tranco-list.eu/} and 1k randomly selected phishing URLs from OpenPhish\footnote{https://openphish.com/} to PhishIntel. 

The results, presented in \hyperref[fig:evaluation]{Figure \ref{fig:evaluation}}, demonstrate that the fast-slow task architecture significantly reduces system latency. Additionally, the blacklist successfully filters out a substantial portion of URLs that do not require further analysis. These findings affirm the overall effectiveness of PhishIntel’s design.
\section{Related Work}
The past few decades have witnessed significant advancements in phishing detection methodologies. Early research primarily focused on blacklist-based\cite{google_safe_browsing} and heuristics-based\cite{shlr} approaches, while more recent studies have shifted towards leveraging machine learning models \cite{google_large_scale}. A prominent framework in this area is RBPD, which leverages computer vision\cite{phishpedia, phishintention, dynaphish} and natural language processing models\cite{knowphish} to analyze webpage information. The booming development of large language models also inspires the utility of LLM agents to extract phishing-related knowledge\cite{phishllm} and understand the webpage content\cite{phishagent}, as a complement to existing RBPDs.

Despite these advancement efforts, to the best of our knowledge, no prior work has addressed the practical deployment of RBPDs. This paper thus presents a novel phishing detection system design, serving as a foundational demonstration to inform and advance the development of more deployment-focused phishing detectors.

% \cite{google_large_scale} leverages a logistic regression classifier with blacklisting for phishing URL detection. With the design goal of low-latency and large-scale detection, the classification workflow is divided into separate processes, each handled by a pool of workers. Tasks are generated for each process as discrete units of work, managed by a task management system. A task queue is implemented for task propagation through the workflow and supports retry mechanisms for failed or timed-out tasks, ensuring no inter-worker coordination.

\section{Conclusion}

In this paper, we introduce PhishIntel, an end-to-end phishing detection system designed for real-world deployment. PhishIntel integrates RBPDs, blacklists, and user reports collaboratively, enhancing both the effectiveness and efficiency of phishing detection. Its fast-slow task system architecture ensures real-time responsiveness and improves user experience. We demonstrate its practical utility through two applications: a phishing intelligence platform and a phishing email detection plugin in Microsoft Outlook. Looking ahead, we plan to optimize PhishIntel for larger-scale deployments.

\begin{acks}
This work was supported in part by the National Research Foundation Singapore, NCS Pte. Ltd. and National University of Singapore under the NUS-NCS Joint Laboratory (Grant A-0008542-00-00).
\end{acks}

\bibliographystyle{ACM-Reference-Format}
\bibliography{reference}

\appendix

\end{document}